\documentclass[aps,prb, twocolumn,footinbib, superscriptaddress,floatfix,longbibliography]{revtex4-2}
\usepackage{amsmath,amssymb} 
\usepackage{bm} 
\usepackage{comment} 
\usepackage{enumitem}
\setlist{noitemsep,leftmargin=*,topsep=0pt,parsep=0pt}
\usepackage{xcolor} 
\usepackage{hyperref}
\hypersetup{
colorlinks=true,
urlcolor= blue,
citecolor=blue,
linkcolor= blue}
\usepackage{graphicx} 

\newcommand{\unit}[1]{\ensuremath{\, \mathrm{#1}}}

\begin{document}

\title{Monte-Carlo ray-tracing studies of multiplexed prismatic graphite analyzers for the cold-neutron triple-axis spectrometer at the High Flux Isotope Reactor}

\author{Adit S. Desai}
\affiliation{School of Physics, Georgia Institute of Technology, Atlanta, Georgia 30332, USA}
\author{Travis J. Williams}
\thanks{Current affiliation: ISIS Neutron and Muon Source, STFC Rutherford Appleton Laboratory, Didcot OX11 0QX, UK}
\affiliation{Neutron Scattering Division, Oak Ridge National Laboratory, Oak Ridge, Tennessee 37831, USA}
\author{Marcus Daum}
\affiliation{School of Physics, Georgia Institute of Technology, Atlanta, Georgia 30332, USA}
\author{Gabriele Sala}
\affiliation{Spallation Neutron Source, Second Target Station, Oak Ridge National Laboratory, Oak Ridge, Tennessee 37831, USA}
\author{Adam A. Aczel}
\affiliation{Neutron Scattering Division, Oak Ridge National Laboratory, Oak Ridge, Tennessee 37831, USA}
\author{Garrett E. Granroth}
\affiliation{Neutron Scattering Division, Oak Ridge National Laboratory, Oak Ridge, Tennessee 37831, USA}
\author{Martin Mourigal}
\affiliation{School of Physics, Georgia Institute of Technology, Atlanta, Georgia 30332, USA}

\date{\today}

\begin{abstract}
A modern cold triple-axis spectrometer to study quantum condensed matter systems is planned for the High Flux Isotope Reactor (HFIR) at Oak Ridge National Laboratory. Here, we describe the conceptual principles and design of a secondary spectrometer using a multiplexed, prismatic analyzer system relying on graphite crystals and inspired by the successful implementation of the Continuous Angle Multiple Energy Analysis (CAMEA) spectrometers at the Paul Scherrer Institute. This project is currently known as MANTA for Multi-Analyzer Neutron Triple-Axis. We report Monte-Carlo ray-tracing simulations on a simple but realistic sample scattering kernel to further illustrate the prismatic analyzer concept’s workings, calibration, and performance. Then, we introduce a new statistical analysis approach based on the prismatic analyzer concept to improve the number of final energies measured on the spectrometer. We also study possible evolutions in the CAMEA design relevant for MANTA.
\end{abstract}

\maketitle

\section{\label{sec:Start} Introduction}

The neutron triple-axis spectrometer (TAS) is an important tool for studying collective excitations of condensed matter systems \cite{brockhouse1995NobelLecture}. In a TAS experiment, the dynamical scattering function of the sample, $S(\mathbf{Q}, E)$, is measured (in selected regions of momentum-energy space) according to the scattering relations for momentum $\mathbf{Q} = \boldsymbol{k}_i - \boldsymbol{k}_f$ and energy transfer $E = E_i - E_f = \frac{\hbar^2}{2 m_n} (k_i^2 - k_f^2)$~\cite{Shirane}. In simplified terms, this requires knowledge of the neutrons' initial and final wave-vectors $\boldsymbol{k}_{i,f}$, the scattering angle $2 \theta$ of the detector, and the azimuthal orientation angle $\phi$ of the sample in the $xy$-scattering plane, which are related by
    \begin{equation}
        Q_x = |k_i| \cos(-\phi) - |k_f| \cos(-\phi + 2 \theta)\, ,
    \end{equation}
    \begin{equation}
        Q_y = |k_i| \sin(-\phi) - |k_f| \sin(-\phi + 2 \theta).
    \end{equation}
In the traditional TAS technique, $S(\mathbf{Q}, E)$ is measured in a point-by-point fashion (within a resolution volume). Measuring the entire sample's response in a given scattering plane typically requires a large number of combinations of $\boldsymbol{k}_i$ ($E_i$), $\boldsymbol{k}_f$ ($E_f$), $2\theta$, and $\phi$, resulting in long measurement times when a survey of $(\mathbf{Q}, E)$-space is desirable. Multiplexing techniques can alleviate this bottleneck by using several analyzers to measure multiple $(\mathbf{Q}, E)$-points simultaneously. Although this can complicate the scattering geometry, the spectrometer's resolution function, and the available background mitigation techniques, this can be advantageous over neutron time-of-flight (TOF) spectroscopy for constrained sample environments, when combined with polarization analysis techniques, to maximize neutron flux using focusing optics, and/or when only a single scattering plane is needed.

Following the distinction made previously \cite{CAMEA_2016}, multiplexing spectrometers can be broadly categorized into two families: wide-angle multiplexing and local multiplexing. In wide-angle multiplexing -- also described as multi-channel detector systems -- several analyzer channels are placed around the sample in a concentric arc, generally with fixed $\boldsymbol{k}_f$~($E_f$). This technique increases the solid angle coverage of the analyzer system but restricts the geometry. One of the earliest examples of this concept is the MADBox spectrometer~\cite{MADBox} at the Forschungsreaktor M\"unchen (FRM), which made use of 61 in-plane analyzer channels. The FlatCone design at the Institut Laue-Langevin (ILL)~\cite{FlatCpone} built upon this concept and uses 31 analyzer channels placed around the sample which scatter out of plane. A particularly efficient example of wide-angle multiplexing is the MACS spectrometer at the National Institute of Standards and Technology (NIST)~\cite{MACS}, which uses analyzers with variable $\boldsymbol{k}_f$ and a combination of filters to suppress higher-order wavelengths. The local multiplexing technique places multiple analyzers on the same channel. An example is the IMPS \cite{IMPS} spectrometer, which places several analyzers at slightly offset angles from each other. The distinct analyzer crystals or blades direct neutrons toward a different spot on a position-sensitive detector (PSD). Similar concepts have been utilized in the SIKA, RITA-II, and PUMA spectrometers \cite{wu2016sika, RITA, PUMA}. A particularly novel design is the HODACA concept, which uses a Rowland ``anti-focusing'' effect  by placing analyzers on a circumferential arc, each of which focuses towards a different detector, resulting in  a constant $E_f$ but large $Q$ coverage \cite{hodaca-zaliznyak}.
\begin{figure*}
    \centering
    \includegraphics[width=0.85\textwidth]{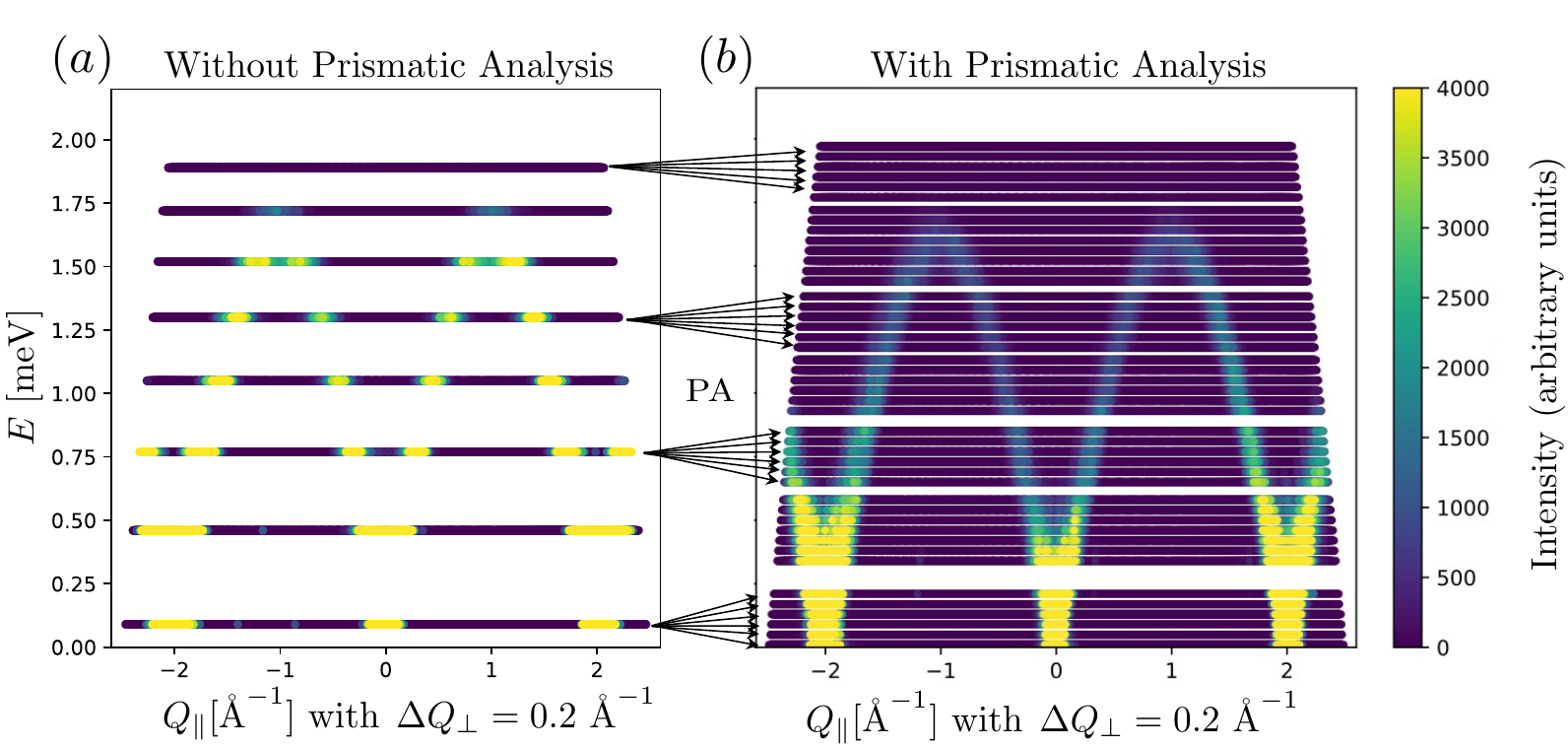}
    \caption{Ray-tracing simulation results illustrating the treatment of inelastic neutron scattering signal for a dispersing excitation of the sample. Simulations utilized $E_i=5.1$~meV and a spectrometer comprising eight analyzer arrays/stations per angular channel (see text for corresponding values of $E_f$). The simulations were treated (a) without and (b) with prismatic analysis (see Sec.~\ref{sec:data analysis} for details on our analysis and calibration technique). The signal is projected and binned in the sample's reciprocal space to produce a momentum-energy slice along $Q_\parallel$ using a perpendicular in-plane integration range of $\Delta Q_\perp = \pm 0.1$~\AA$^{-1}$ around the average $Q_\perp=-2.0{\rm \AA}^{-1}$. $Q_\parallel$ and $Q_\perp$ refer to the momentum transfer along and orthogonal to, respectively, the viewing momentum-energy slice. The prismatic analysis technique resolves the signal by accounting for the distribution in final neutron energy for different pixels along a position-sensitive detector.}
    \label{fig:teaser}
\end{figure*}
One of the more recent developments in multiplexing TAS design is the Continuous Angle Multiple Energy Analysis (CAMEA) spectrometer, which is installed and commissioned at the Paul Scherrer Institute (PSI) \cite{CAMEA_2016, CAMEA_2020, CAMEA_2023}. CAMEA uses the nearly perfect transmission of highly-oriented pyrolytic graphite (HOPG) crystals \cite{HOPG} to combine the local and wide-angle multiplexing techniques. While traditional local multiplexing (e.g. IMPS \cite{IMPS}) requires analyzers to be placed at slightly different scattering angles, CAMEA places analyzers directly behind each other, simplifying the geometry. Each of the eight (8) arrays of analyzers, or stations, focuses a band of different $\boldsymbol{k}_f$~($E_f$) out of the scattering plane at specific locations on a position-sensitive detector, hence creating a quasi-energy-resolved detector. Eight (8) channels of this multi-analyzer system are placed at different $2\theta$ angles, resulting in wide-angle coverage. A similar concept has been successfully implemented at spectrometers like MultiFlexx \cite{MultiFlEXX} and BAMBUS \cite{BAMBUS} and it is planned for BIFROST at the European Spallation Source (ESS) \cite{BIFROST}. 

Maintenance and upgrade plans for the High Flux Isotope Reactor (HFIR) at Oak Ridge National Laboratory~\cite{HBRR2023} offer the opportunity to design and execute the construction of a state-of-the-art cold-neutron triple-axis spectrometer on a new cold guide-hall location, NB-6. In this paper, we employ Monte-Carlo ray tracing simulations to explore the concept and the performance of a secondary spectrometer relying on the CAMEA design and an optimized primary spectrometer. The design principles, optimization, and performance of the primary spectrometer (guide system, velocity selector, and monochromator) are discussed in Ref.~\cite{GRANROTH2024169440}. The secondary spectrometer is currently known as the Multi-Analyzer Neutron Triple Axis (MANTA) spectrometer. This project offers the opportunity to explore and implement various forms of prismatic analysis strategies, which greatly enhances the resolving power of the spectrometer, as shown conceptually in Fig.~\ref{fig:teaser}.

This paper is organized as follows. In Sec.~\ref{sec:Design}, we outline the geometrical and technical parameters of our model instrument, as well as the multiplexing and prismatic analysis concepts it utilizes. In Sec.~\ref{sec:Methods}, we detail the methods of our ray-tracing Monte-Carlo simulations on a realistic sample kernel to benchmark the performance of our model instrument. In Sec.~\ref{sec:data analysis} we analyze the simulation data and present the principles of prismatic analysis along with the demonstration of a novel approach called Positionally-Calibrated Prismatic Analysis (PCPA). In Sec.~\ref{sec:manta-1}, we present the results of our ray-tracing simulations and data analysis approach on a realistic sample kernel and present how modifications to the original design impact data acquisition and quality. We conclude this work in Sec.~\ref{sec:Conclusion}.

\begin{figure}
    \includegraphics[width=0.99\columnwidth]{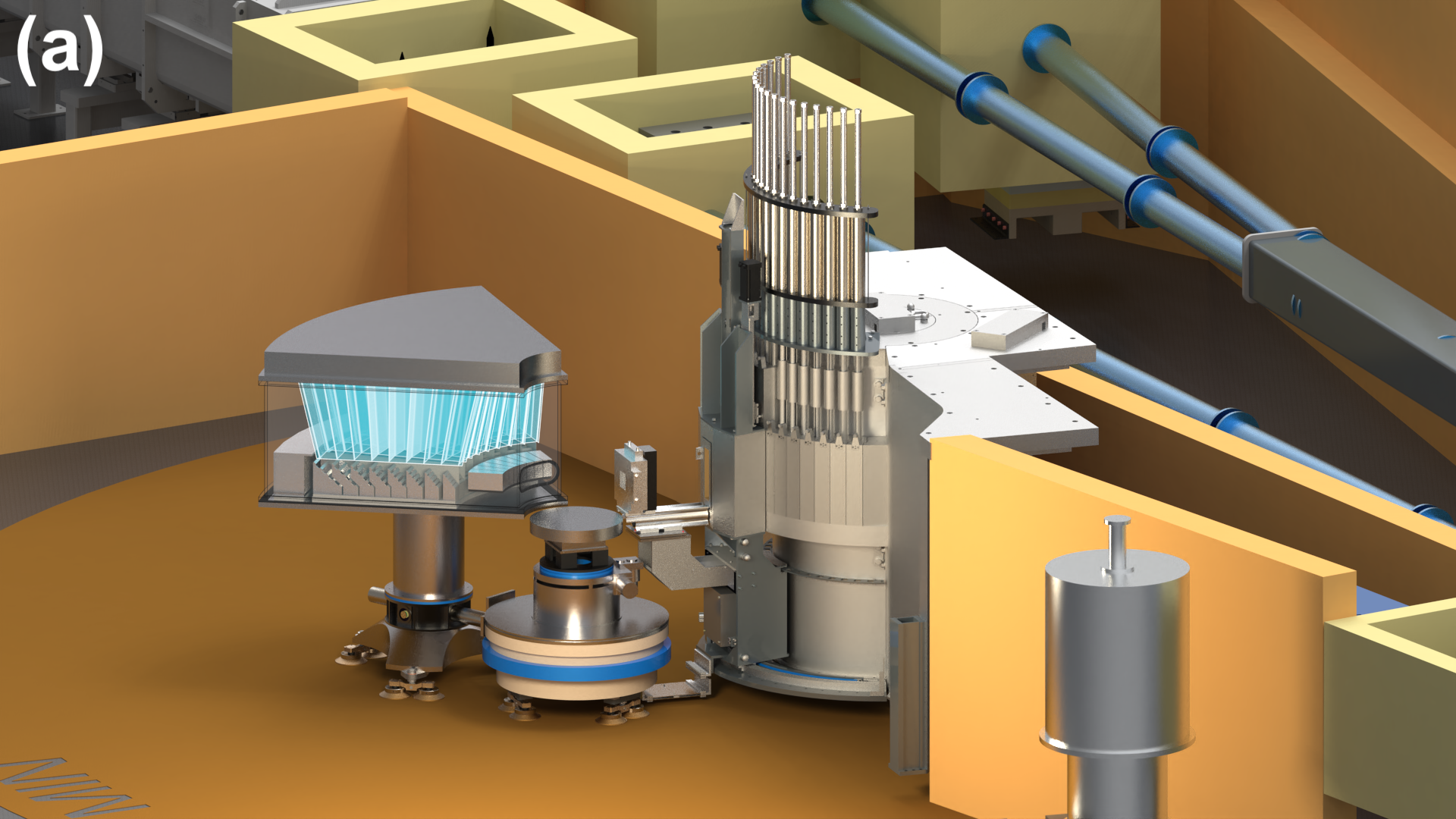}
    \includegraphics[width=0.99\columnwidth]{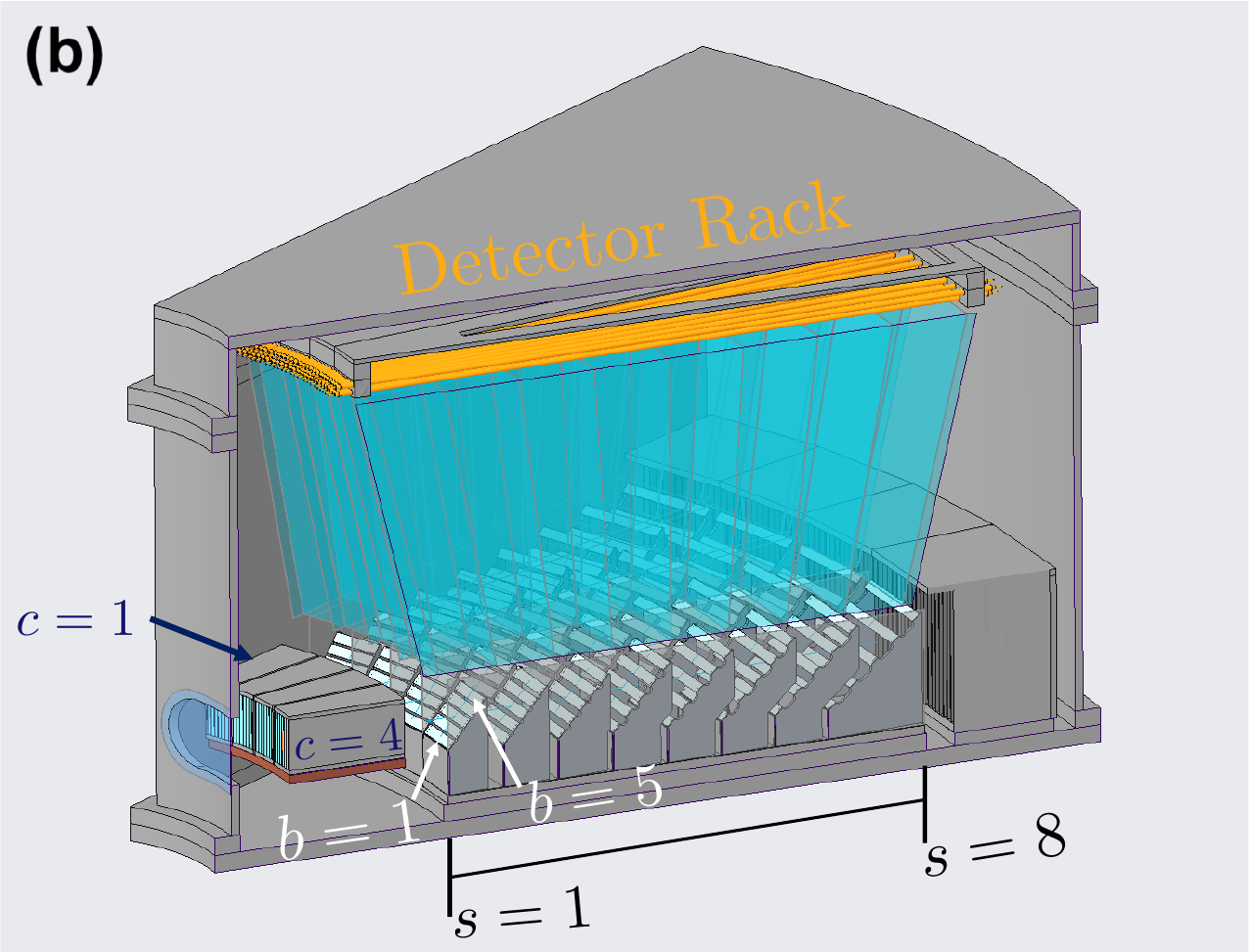}
    \caption{Preliminary conceptual engineering renderings for the MANTA spectrometer at Oak Ridge National Laboratory's High Flux Isotope Reactor. (a) Overview of the spectrometer on a newly designed NB-6 neutron guide system in the reconfigured cold neutron guide hall of HFIR. (b) Close-up view of a possible implementation of the secondary spectrometer using the CAMEA \cite{CAMEA_2016, CAMEA_2023} design showing $c=1-$ through $c=4$ (of eight total) angular channels, $N_s=8$ analyzer stations per channel, and $N_b=5$ HOPG blades per station.}
    \label{fig:MANTA-Concept}
\end{figure}

\section{\label{sec:Design} Instrument Concept and Model}

The model of the primary spectrometer comprises an optimized neutron guide system with supermirror coatings that directs neutrons from the HFIR cold-neutron source (with the parameters of the existing HFIR cold-source) onto a 300~mm $\times$ 168~mm doubly focusing monochromator. The primary was optimized for incident energies of 2.4 to 20 meV, see Ref.~\cite{GRANROTH2024169440} for details. The sample is located 160~cm from the monochromator. An aspirational engineering rendering of the spectrometer in the HFIR cold guide hall is presented in Fig.~\ref{fig:MANTA-Concept}(a). Using a focusing monochromator increases the neutron flux on the sample. However, it yields a broad divergence of the incident beam at the sample position, affecting the spectrometer's resolution. By incorporating the guide system and the doubly focusing monochromator in our simulations (see Section \ref{sec:Methods}), these resolution effects can be accounted for. In some instances in Section~\ref{sec:manta-1}, we utilized a simplified, ``toy model" design of the full primary spectrometer. The ``toy model" was designed to ease computational intensity and used an idealized virtual source and curved monochromator that produced a similar spot size and energy resolution at the sample location.

For the secondary spectrometer, we adapted the back-end design of CAMEA \cite{CAMEA_2016} to our virtual instrument, see Fig.~\ref{fig:MANTA-Concept}(b) for an engineering diagram. As the planned use of a beryllium filter in the scattered beam requires analyzed energies to below 5.2 meV, and geometrical constraints require the analyzed energy to be above 2.4 meV, the CAMEA design is particularly well suited for our purposes. The design consists of $N_c=8$ angular channels, each of which have $N_d=13$ detectors that span $\Delta(2\theta)\!=\!7.5^{\circ}$ per channel, see Fig.~\ref{fig:MANTA-Concept}(b). Each channel comprises $N_s$ analyzer stations in series, each built from $N_b$ blades of flat, highly-oriented pyrolytic graphite (HOPG) crystals. The analyzer blades of a given station have a mosaic of 1$^\circ$, equivalent to 60 arc minutes (60'), and are rotated to fulfill the Bragg condition for a given $E_f$. We label the analyzer station with index $s=1\cdots N_s$. Furthermore, analyzer stations between different channels are focused using Rowland geometry \cite{Shirane, Skoulatos_2012}. Analyzer stations placed further from the sample are larger to account for decreased solid angle coverage for a given crystal size with increasing distance. Additional details of the instrument model are provided in Tab. \ref{table:properties}. In some instances in Section~\ref{sec:manta-1}, we modify the CAMEA design by changing the number of stations $N_s$ and the analyzer mosaic.
    
\begin{table*}
\begin{ruledtabular}
\begin{tabular}{c c c c c c c c c}
    \textbf{Analyzer Station \#, $s=$} & \textbf{1} & \textbf{2} & \textbf{3} & \textbf{4} & \textbf{5} & \textbf{6} & \textbf{7} & \textbf{8}\\ 
    \hline
    Energy $E_f(s)$ (meV) &  3.21 & 3.38 & 3.58 & 3.8 & 4.05 & 4.33 & 4.64 & 5.01\\
    \hline
    Sample to Analyzer Distance (mm) & 939 & 994 & 1057 & 1120 & 1183 & 1246 & 1310 & 1375 \\
    \hline
    Analyzer to Detector Distance (mm) & 707 & 702 & 700 & 701 & 703 & 709 & 717 & 727 \\ 
    \hline
    HOPG Segment Length (mm) & 72 & 82 & 92 & 103 & 113 & 120 & 129 & 140 \\
    \hline
    HOPG Segment Height (mm) & 12.0 & 12.0 & 12.5 & 13.0 & 13.5 & 14.0 & 15.0 & 16.0 \\
\end{tabular}
\end{ruledtabular}
\caption{The parameters of each analyzer station in a single angular channel as adapted from the CAMEA design~\cite{CAMEA_2016,CAMEA_2020,CAMEA_2023}. The area of HOPG crystal blades of a given station increases with $E_f$ to account for the decrease in solid angle coverage at larger sample-analyzer distances.}
\label{table:properties}
\end{table*}

\begin{figure}[t!]
    \centering
    \includegraphics[width=0.49\textwidth]{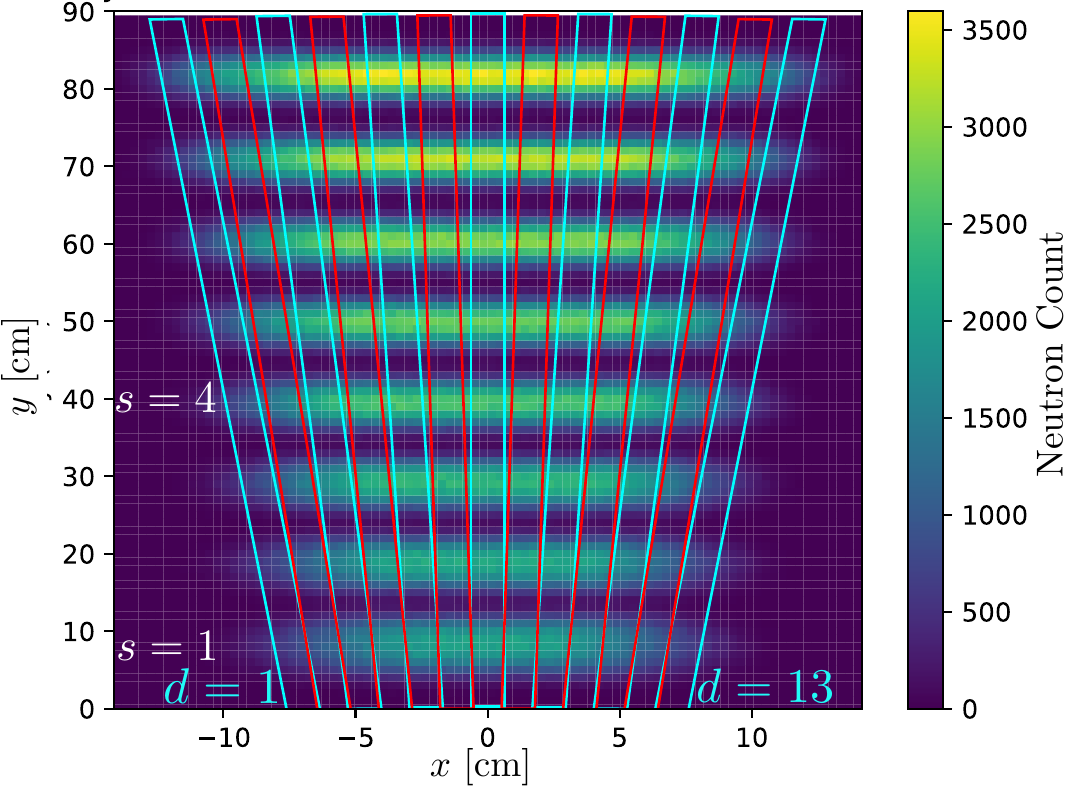}
    \caption{Neutron ray-tracing simulations of a single channel with $N_s=8$ stations of $N_b=5$ for the parameters of Tab. \ref{table:properties} to illustrate the local multiplexing effect. A collimated ``white'' beam of neutrons with uniform spectral density between {$E_i=1.0$~meV and $7.0$~meV} is directed at the secondary spectrometer. For illustration, instead of realistic PSD tubes used later in this work, neutron counts are collected here using a rectangular PSD detector above the analyzer stations. The lower rack (cyan) and upper rack (red) of the detectors are outlined to show their coverage.}
    \label{fig:psd}
\end{figure}

The analyzer blades reflect neutrons from the sample to an out-of-scattering plane direction and towards a series of horizontal position-sensitive detectors (PSDs), laid out in a radial pattern above the stations. Furthermore, the radial pattern has two vertical rows to minimize gaps between detectors.
Each detector has an active length of $\approx0.9$~m and a diameter of 12.6mm. By focusing neutrons on specific locations on the detectors, each analyzer station generates a ``strip" of neutrons associated with a particular $E_f$. As illustrated in Fig.~\ref{fig:psd} for the parameters of Tab. \ref{table:properties}, this approach achieves local multiplexing by avoiding overlap between strips generated by different stations. Thus, at minima, this allows the acquisition of $N_s=8$ $E_f$ values for each channel.

\begin{figure}
    \centering
    \includegraphics[width=0.49\textwidth]{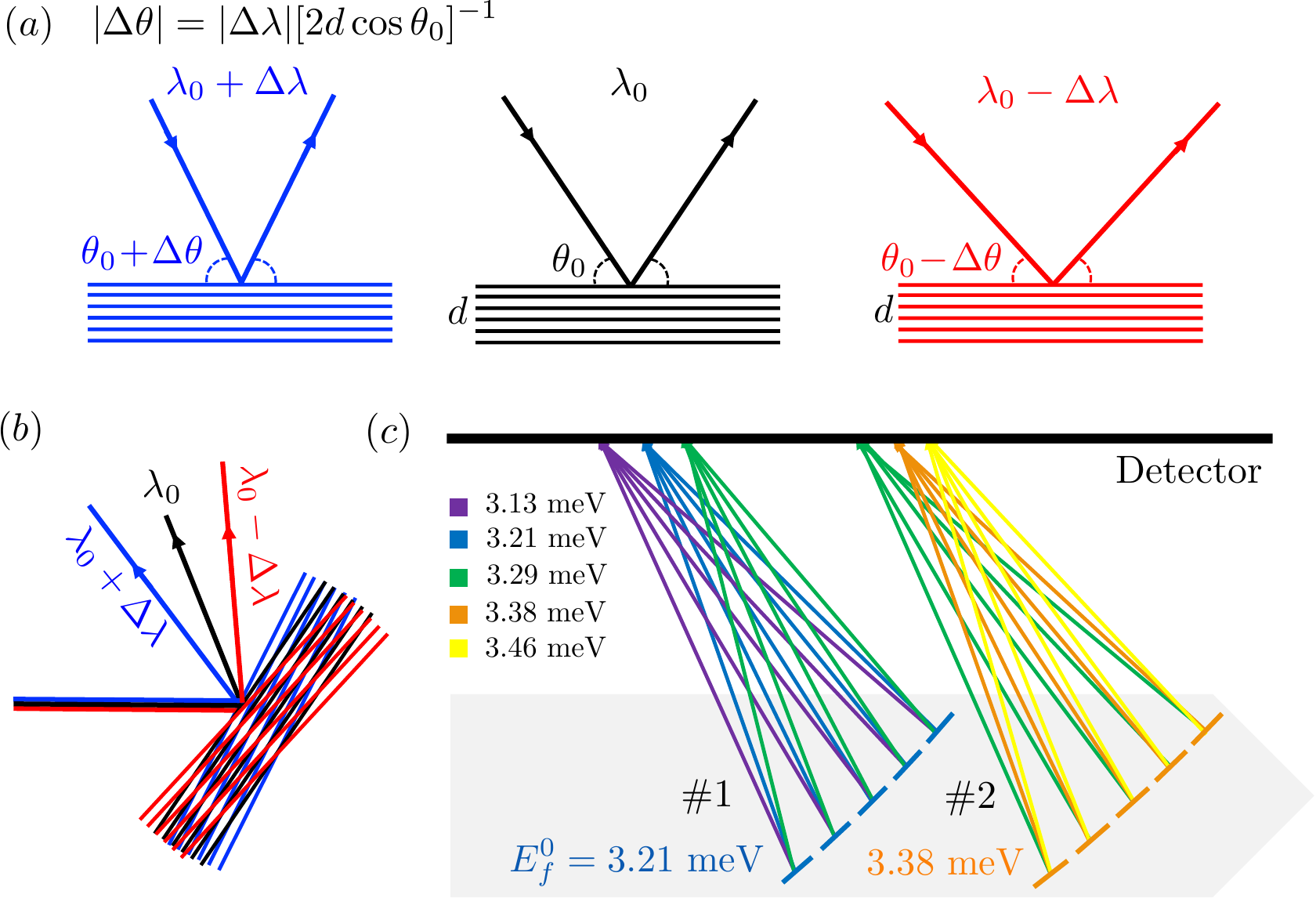}
    \caption{((a) Relation between small angular and wavelength deviations for a perfect analyzer blade. (b) In prismatic analysis, neutrons with proximate wavelengths/energies experience different angular deviations from fulfilling the Bragg scattering condition. (c) Using a position-sensitive detector far above the analyzer stations, different wavelengths become discernible on the detector. Note that in this diagram the 3.29~meV neutrons are scattered by two analyzers, an effect also captured in our ray-tracing simulations. 
    }
    \label{fig:prismatic}
\end{figure}

Though this method of discrete angular and analyzer channels to form a multiplexing instrument has been accepted as a viable approach for decades, one of the advantages and innovations of the CAMEA concept is the utilization of the prismatic analyzer concept~\cite{Birk_prismatic, CAMEA_2016, CAMEA_2020}. The foundations of prismatic analysis come from a first-order expansion of Bragg's law from a finite mosaic analyzer crystal (with lattice spacing $d$). This yields a linear relationship between small deviations in incident angles $\Delta \theta$ and small deviations in reflected neutron wavelength $\Delta \lambda$ around a nominal Bragg reflection ($\sin\theta_0=\lambda_0/2d$), as illustrated in Fig.~\ref{fig:prismatic}(a). As a result of analyzer mosaic, neutrons with proximate wavelengths from the nominal wavelength $\lambda_0$ are Bragg-scattered in different directions, Fig.~\ref{fig:prismatic}(b). Using a position-sensitive detector, this effect can be advantageously used to encode small wavelength deviations into detector positions, Fig.~\ref{fig:prismatic}(c). Since its inception,  prismatic analysis has been featured in the design of the planned time-of-flight instruments BIFROST \cite{BIFROST} and FARO \cite{FARO}.

\section{\label{sec:Methods} Simulations: Methods and Sample Kernels}

To assess the performance and validate the concept of our spectrometer model, we employ neutron ray-tracing simulations using the McStas package~\cite{McStas}. McStas is a state-of-the-art tool used in designing modern neutron scattering instruments \cite{BIFROST, FARO, BAMBUS, IMPS, MACS, MultiFlEXX, CAMEA_2016, RITA, CHESS, CHESS2, AMETERAS, ARCS, Wish}. The program uses Monte Carlo techniques to simulate the behavior of neutron scattering instrument components with high accuracy. It also allows testing the instrument's performance on various simple sample scattering kernels to mimic actual experiments. In this work, we employed McStas versions 2.7, 3.1, and 3.4.

The primary spectrometer, from the cold source to 15~cm before the sample, was simulated in McStas for various choices of incident energy $E_i$, using the design of Ref.~\cite{GRANROTH2024169440}. These results were pre-compiled to form a set of simulated incoming beam profiles for the secondary spectrometer. The latter was simulated by adapting the McStas code of the PSI-CAMEA back-end, which was generously provided to us by its designers \cite{CAMEA_2016}. The virtual instrument comprises $N_b=5$ analyzer blades per station, $N_s=8$ stations per channel,  $N_c=8$ channels with $N_d=13$ detectors per channel. This results in a total continuous scattering angle coverage of $8\times7.5^\circ\!=\!60^\circ$ and a total of  $N_s * N_c * N_d = 832$ points in $S(\vec{Q}, \omega)$ measured simultaneously. While the actual spectrometer uses filters and a radial collimator after the sample, we did not include these components in our simulations to maintain a continuous coverage; including collimators in our simulations is left for future work.  A 10 $\times$ 10~cm rectangular slit simulates the opening in front of the analyzer stations, allowing full illumination of the analyzers and unrestricted angular coverage. Unless otherwise noted, we used an infinitely thin HOPG analyzer with a mosaic of 60'. Note that modeling the analyzer in this manner does not take into account the background signal that arises from thermal diffuse scattering and phonons in HOPG, nor the effect of the decrease in peak reflectivity as a function of crystal mosaic \cite{HOPG, HOPG_Petersen, Shirane}.

We employed three types of sample kernels. The first virtual sample was a full cylinder of polycrystalline vanadium with a 2~cm height and a 1.0~cm radius. We used this incoherent scattering kernel to obtain calibration runs (see Section \ref{sec:data analysis}) of the analyzer system and the position-sensitive detectors. Then, a virtual single crystal with a similar cylindrical geometry was prepared using the McStas phonon kernel component to mimic a simple bosonic excitation (acoustic magnon or phonon) dispersing in three dimensions with a bandwidth below 2~meV and uniform intensity across the Brillouin zone. The parameters of this inelastic scattering kernel are found in Tab.~\ref{tab:sample-parameters}. Finally, we made use of a single crystal elastic Bragg scattering sample with a mosaic of 10', lattice constant c = 6.71 $\unit{\AA}$, and the same geometry as the inelastic scattering kernel, which produced a single Bragg peak at $\vec{Q} = (0,0,2)$ ($|\vec{Q}| = 1.85 \unit{\AA^{-1}}$). This sample was used to calculate the elastic $\vec{Q}$ broadening due to the instrument.

\begin{table}[]
    \centering
    \begin{tabular}{|c c|}
    \hline
        \textbf{\centering Sample Kernel Parameters} & \\
        \hline
         radius (cm) & 1.0 \\
        \hline
         height (cm) & 2.0\\
         \hline
         $\sigma_{\rm abs}$ (barns) & 0.171 \\
         \hline
         $\sigma_{\rm inc}$ (barns) & 0.32 \\
         \hline
         fcc Lattice Constant (${\rm \AA}$) & 6.28 \\
         \hline
         Scattering Length (fm) & 9.405\\
         \hline
         Mass (amu) & 207.2 \\
         \hline
         Speed of Sound ${\rm meV \AA^{-1}}$ & 2.5 \\
         \hline
         Temperature (K) & 10 \\
         \hline
         Debye-Waller Factor & 1 \\
         \hline
    \end{tabular}
    \caption{Parameters for the McStas \cite{McStas} phonon kernel used in our simulations.}
    \label{tab:sample-parameters}
\end{table}

The simulations with the inelastic scattering crystal kernel were conducted to model a realistic scattering experiment covering energy transfers between $E=0$~meV and $E=2$~meV. As the detector system has an overall $2\theta$ coverage of $60^\circ$, the single crystal sample was simulated for several positions of the detector system. Using the lowest angle of the detector systems as a reference, we aimed to measure 5 $2\theta_i$ values centered around three reference points, $6^\circ$, $66^\circ$, and $116^\circ$. However to minimize scattering signal from non-interacting neutrons, simulations where the lowest detector angle was below $2.5^\circ$ or the largest detector angle was at or above $180^\circ$ were not simulated, leading to a net total of 12 positions. For each detector position, the single crystal sample was rotated in steps of $\Delta\phi = 2.5^{\circ}$ and we ran the Monte-Carlo process with $>10^8$ neutrons for each sample orientation. This was repeated for two $E_i$ values, 5.1 meV and 5.27 meV, to produce the simulated results. The kernel was designed to measure the entire dispersion curve with $E_i = 5.1$ meV, and the second incident energy of $E_i = 5.27$ meV was selected to fill in the gaps in energy space. These results are shown in Section \ref{sec:manta-1}.

Additional simulations with an elastic Bragg scattering sample were conducted to determine the elastic $\vec{Q}$ broadening as a function of $E_i$. To do this, the full primary and secondary spectrometer were used in a similar setup as the inelastic sample simulations. 
However, only one angular channel was used and a single slit with an opening of $1^\circ$ was placed in half the simulations. Using the known position of the Bragg peak as a reference, $2\theta$ was varied from a range of $\Delta 2\theta = \pm 12^\circ$ from the Bragg peak in steps of $1^\circ$ and sample was rotated in a range of $\Delta \phi = \pm 7$ in steps of $0.5^\circ$. This was repeated for three $E_i$ values, 3.21 meV, 4.05 meV, and 5.01 meV. The results are tabulated in Section \ref{sec:manta-1}.

\section{\label{sec:data analysis} Prismatic Analysis and Data Processing} 


For a given analyzer channel and orientation of the sample, it is possible to use prismatic analysis to obtain the scattering function $S(\mathbf{Q}, E)$ for significantly more values of energies than the number of stations $N_s$ as shown conceptually on Fig.~\ref{fig:teaser}. While prismatic analysis relies on post-measurement data analysis techniques that require careful calibration, it is possible to devise several such strategies. 

CAMEA has successfully implemented an ``energy-calibration" approach \cite{MJOLNIR, CAMEA_2020, CAMEA_2023}. As shown in Fig.~\ref{fig:psd}, the neutrons scattered by a given analyzer station -- nominally associated with $E_f(s)$ -- produce a well-separated strip (or analyzer-specific area) on the position-sensitive detector. Each strip in the detector space is broken into $N_p$ prismatic sub-regions (with $N_p$ ranging from $3$ to $7$). The integrated signal detected in a given sub-region is entirely associated with a particular $E_f(s,p)$, proximate, but not equal to the reference $E_f(s)$, which effectively yields a continuous energy coverage but requires careful calibration~\cite{CAMEA_2023}. To determine the average $\bar{E}_f(s,p)$ analyzed by the $p$-th detector sub-region associated with analyzer station $s$, a calibration is performed by measuring the scattering from a vanadium sample as a function of small steps in $E_i$ ($\Delta E_i \approx 0.01$ meV). The incoherent vanadium signal integrated over a given sub-region yields a Gaussian profile as a function of energy, which can be fitted to obtain $\bar{E}_f(s,p)$, hence the name of Energy-Calibrated Prismatic Analysis (ECPA). Because the efficiency of the prismatic analysis across sub-regions is not uniform, this approach also conveniently yields an intensity normalization $A(s,p)$ to take analyzer transmission and prismatic/detector efficiency into account. 

\begin{figure}
    \centering
    \includegraphics[width=0.47\textwidth]{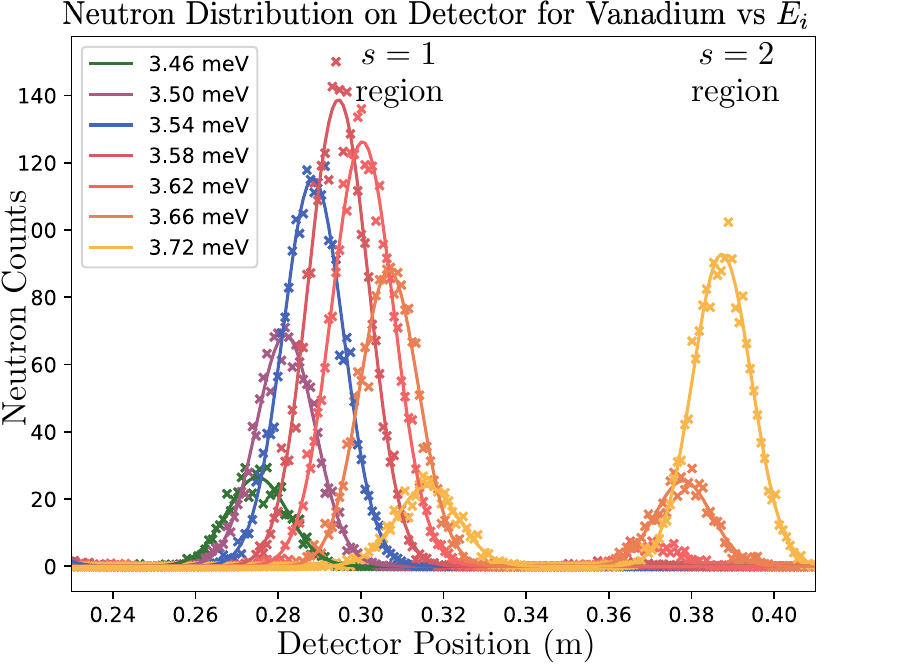}
    \caption{Intensity versus detector position for several calibration runs with vanadium as a sample. The distribution of neutron counts for given incident energies $E_i$ (indicated on the graph) are over-plotted. For each $E_i$, the neutron count distribution is fitted to a Gaussian profile. For values of $E_i$ far from the nominal $E_f$, the Bragg condition may be fulfilled by multiple stations, as seen here through the additional peaks in the detector region associated with the $E_f(s=4)=3.80$~meV station.}
    \label{fig:pcpa-calibration}
\end{figure}

While the above approach works very well, it is interesting to compare it to an alternative method of data analysis which we call the Positionally-Calibrated Prismatic Analysis (PCPA) technique. As opposed to defining discrete sub-regions in detector space and calibrating the average energy being detected, the alternative approach defines discrete steps for the final neutron energies $E_f$ and reconstructs the spatial distribution of each $E_f$ on the detector system. This approach is demonstrated in Fig.~\ref{fig:pcpa-calibration}. The rationale for this approach is that neutron detection along the detector is very well-defined by the detector electronics, which for the CAMEA implementation at PSI has 1024 pixels over an active length of $\approx0.9$~m \cite{CAMEA_2020}. Just as with ECPA, calibration with a vanadium standard is essential.

To calibrate the analyzer-detector system, the primary spectrometer is tuned to a particular incident energy ($E_i$) to analyze the elastic incoherent scattering from the vanadium kernel. It is important to note that the vanadium calibration procedure must use a vanadium sample of the same height as the sample to be measured to properly account for the correct vertical divergence in the beam after scattering. We have verified that the $2\theta$-position of the secondary spectrometer does not affect the results; thus, the calibration procedure is performed at a single position of the detector system. After scattering from the vanadium and a given analyzer blade, neutrons with energy $E_f=E_i$ are detected at particular locations on the PSD. This yields a Gaussian distribution for each $E_f$, as seen in Fig.~\ref{fig:pcpa-calibration}. This process is repeated by sweeping $E_i$ across various bands of energy centered around the nominal $E_f(s)$ of the analyzers, that is  $E_i = E_f(s) \pm 0.12$~meV is scanned in steps of $0.04$~meV for $s=1\cdots N_s$ (total of $\approx 7$ $E_i$'s per station $s$).  
\begin{figure}
    \centering
    \includegraphics[width=0.49\textwidth]{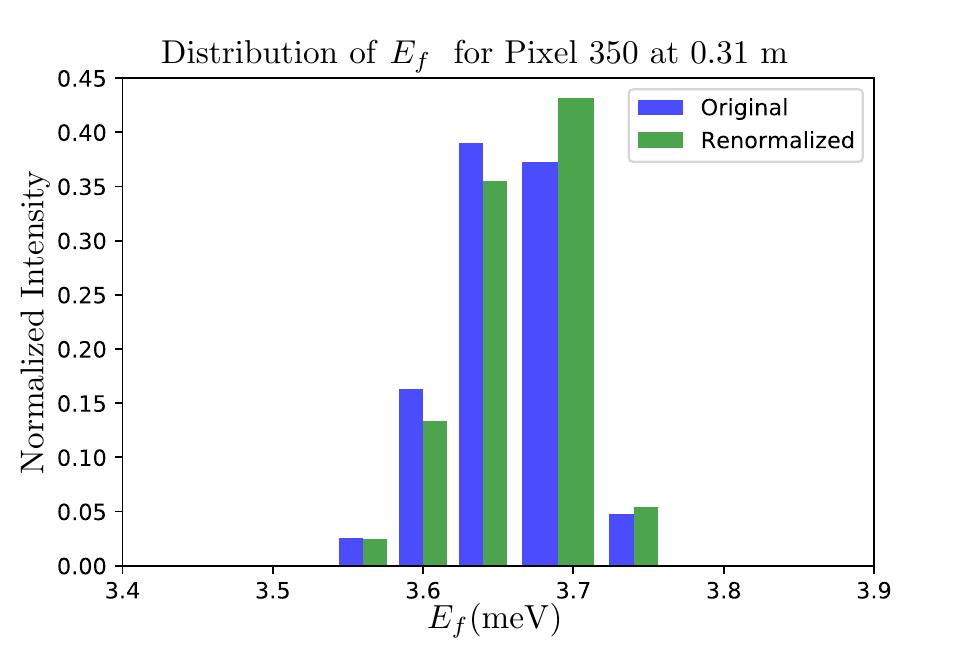}
    \caption{For each pixel, our calibration procedure yields a histogram distribution representing the likelihood that a neutron detected on a given pixel had energy $E_f$. Here, results are shown for a representative pixel. Results with and without scaling to account for prismatic efficiency are shown.}
    \label{fig:histograms}
\end{figure}

This process provides the distribution of neutron counts versus detector position for each discrete $E_f=E_i$ within the calibration set. When comparing the different $E_f$'s, it is clear that the total detector-integrated intensities $I(E_f)$ vary significantly due to the radial arrangement of the detectors and the fact that the additional energies measured from the prismatic concept are intrinsically less intense, see Fig.~\ref{fig:pcpa-calibration}. To correct for this effect, we introduce a scale factor, $G(E_f) = I(E_f=3.21 \unit{meV})/I(E_f)$, where $I(E_f)$ is the integrated intensity of a given $E_f$ over the entire detector, therefore summing over all Gaussian profiles ascribed to a specific $E_f$. The integrated intensity of the $E_f=3.21$~meV energy was arbitrarily chosen as a reference as it is the most intense. The scale factor is then applied as a prefactor to each Gaussian profile in the detector space. The last, and central step is to transpose the ensemble of these scaled distributions (of neutron counts versus position at fixed $E_f$) into a new set of distributions representing neutron counts versus $E_f$ at fixed position. In other words, we create normalized histograms of final energies \textit{for each pixel}, representing the likelihood that a neutron detected on a given pixel had energy $E_f$. A representative result of such a histogram is shown in Fig. \ref{fig:histograms}. Note that the scaling factor $G(E_f)$ only affects the relative distribution of energies within a histogram; in the end, the scaling does not affect the net flux at the detector. Additionally, to mimic the effect of cross-talk baffles, neutrons that land in between strips of neutrons (as seen in Figure \ref{fig:psd}) are discarded. In the next section, we employ the PCPA technique on a realistic sample kernel for a virtual spectrometer to assess its feasibility.

\section{\label{sec:manta-1} Results and Secondary Spectrometer Optimization}

\begin{figure}
    \centering
        \includegraphics[width =0.40\textwidth]{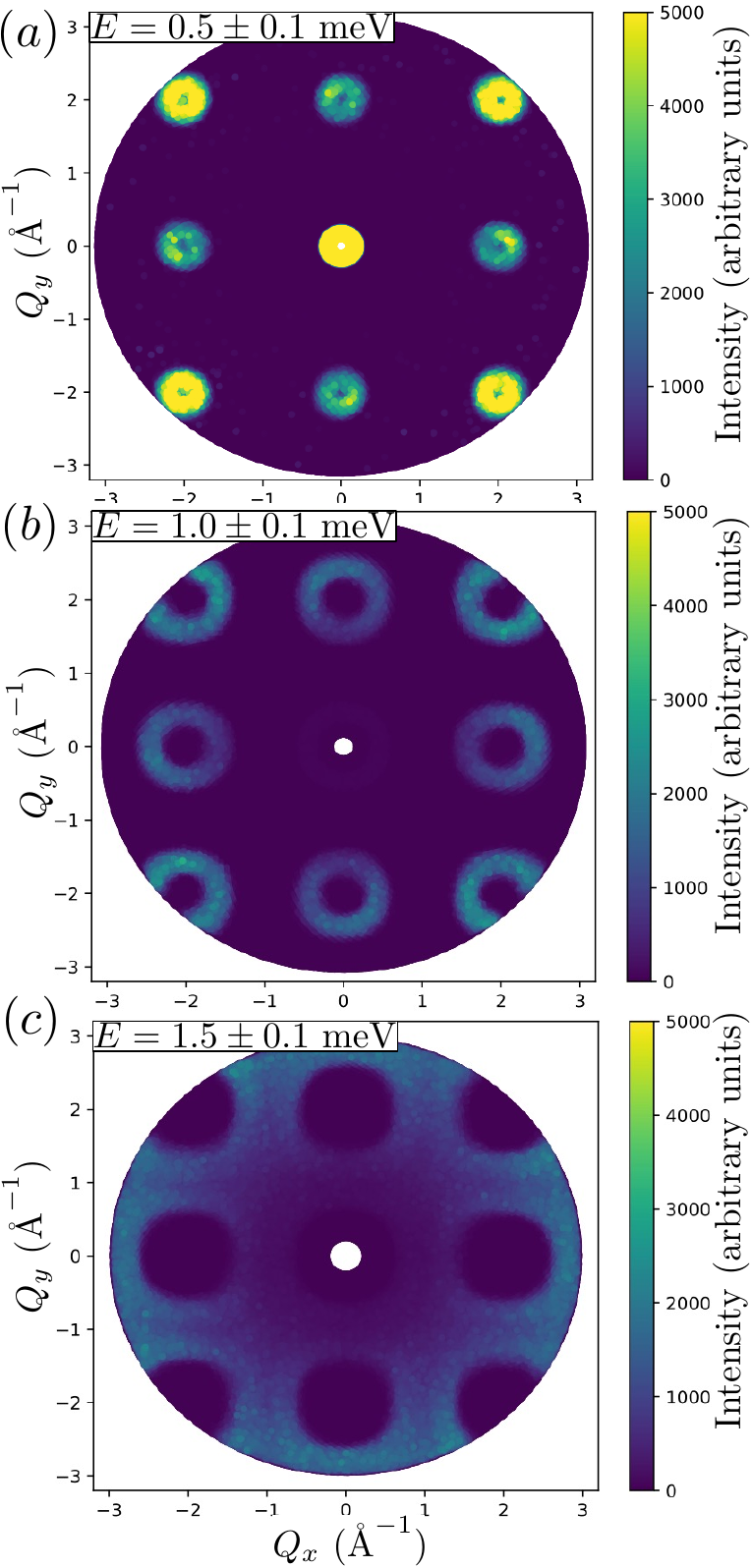}
    \caption{Scattering intensity represented as constant energy slices for the results of our single-crystal simulations with (a) $E=0.5 \pm 0.1$~meV, (b) $E=1.0 \pm 0.1$~meV, and (c) $E= 1.5 \pm 0.1$~meV. These simulations utilize full models for the primary spectrometer and secondary spectrometer as described in Sec.~\ref{sec:Design}.  Two incident energies, several positions of the detector system, and full rotations of the sample were used to acquire the data which was subsequently analyzed using the Positionally-Calibrated Prismatic Analysis technique described in Sec.~\ref{sec:data analysis}.}
    \label{fig:eSlices}
\end{figure}

\begin{figure}
    \includegraphics[width=0.45\textwidth]{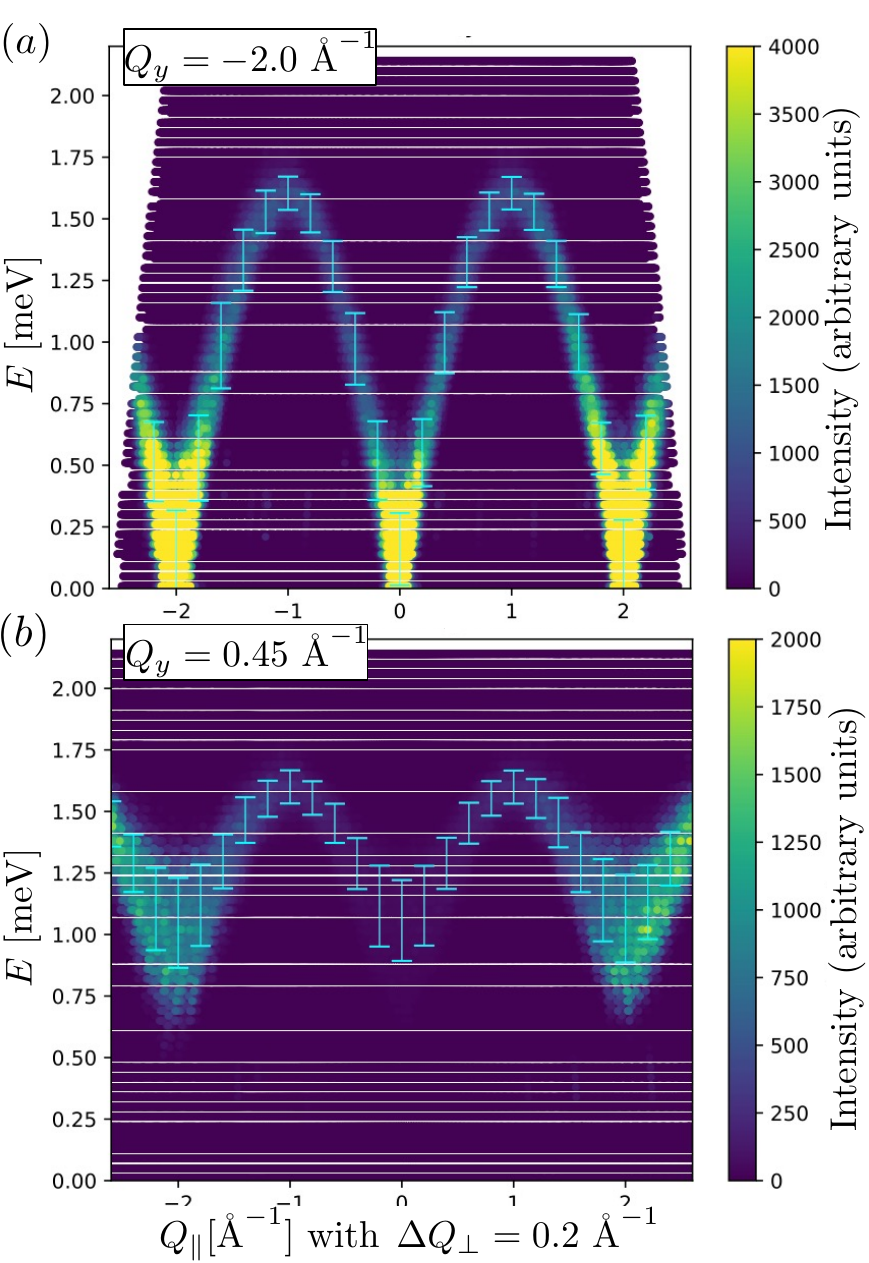}
    \caption{Scattering intensity represented as momentum-energy slices for the results of our single-crystal simulations with the momentum transfer $Q_x$ as the scan direction $Q_\parallel$ and $Q_y$ as $Q_\perp$ direction integrated over $\pm 0.1$~{\AA}$^{-1}$ for: (a) $Q_y=-2.0$~\AA\ and (b) $Q_y=+0.45$~\AA$^{-1}$. Two incident energies, several positions of the detector system, and full rotations of the sample were used to acquire the data which was subsequently analyzed using the Positionally-Calibrated Prismatic Analysis technique described in Sec.~\ref{sec:data analysis}.
    }
    \label{fig:qyslices}
\end{figure}

\subsection{Original Model}

We now turn to our results of the expected performance of the full MANTA model discussed in Sec.~\ref{sec:Design} using the single-crystal kernel and the simulations methods of Sec.~\ref{sec:Methods} and the PCPA calibration technique discussed in Sec.~\ref{sec:data analysis}. In Fig.~\ref{fig:eSlices}, we present the results for a complete set of single-crystal rotations as constant energy-slices representing the neutron scattering intensity in the $(Q_x,Q_y,0)$ scattering plane for three different energy transfers spanning the bottom, middle, and top of the dispersion band. In Fig.~\ref{fig:qyslices}, we present momentum-energy slices across the same datasets. These results highlight the efficiency of the secondary spectrometer in capturing the entire volume of $S(Q_x,Q_y,E)$ with only two incident energies $E_i$. 
Furthermore, it reveals resolution effects that are expected on triple-axis spectrometers\cite{Shirane}, namely focusing and de-focusing effects, see e.g. the distribution of intensity on excitation rings in Fig.~\ref{fig:eSlices}(a) and Fig.~\ref{fig:eSlices}(b), as well as energy-broadening due to momentum integration and analyzer mosaic, see e.g. the width of the dispersion (cyan) in Fig~\ref{fig:qyslices}. The width is calculated in intervals of $\Delta Q_x = 0.2\unit{\AA^{-1}}$ as the full-width-half-maximum (FWHM) of the intensity integrated over a width of $Q_x \pm 0.05$.

\subsection{Studies of Analyzer Mosaic Effects}

By its nature, prismatic analysis, particularly the PCPA technique we introduce, invites exploring changes to the original instrument model to measure more final energies simultaneously. In the following two subsections, we study two possible modifications of the instrument hardware: (1) changing the analyzer mosaic and (2) changing the number of analyzer stations. Analyzer crystal mosaic is an important parameter in determining the resolution of a triple-axis spectrometer: the larger the crystal mosaic, the larger the bandwidth of energies scattered by the analyzer. In a traditional setup, analyzers with a large mosaic ($>1^\circ$) result in a poorer energy resolution. However, the foundational principle of the prismatic analysis is that the energy resolution can be independent of crystal mosaic through the use of distance collimation \cite{Birk_prismatic, marko2014calculations, CAMEA_2016}. 

\begin{figure}
    \centering
    \includegraphics[width=0.45\textwidth]{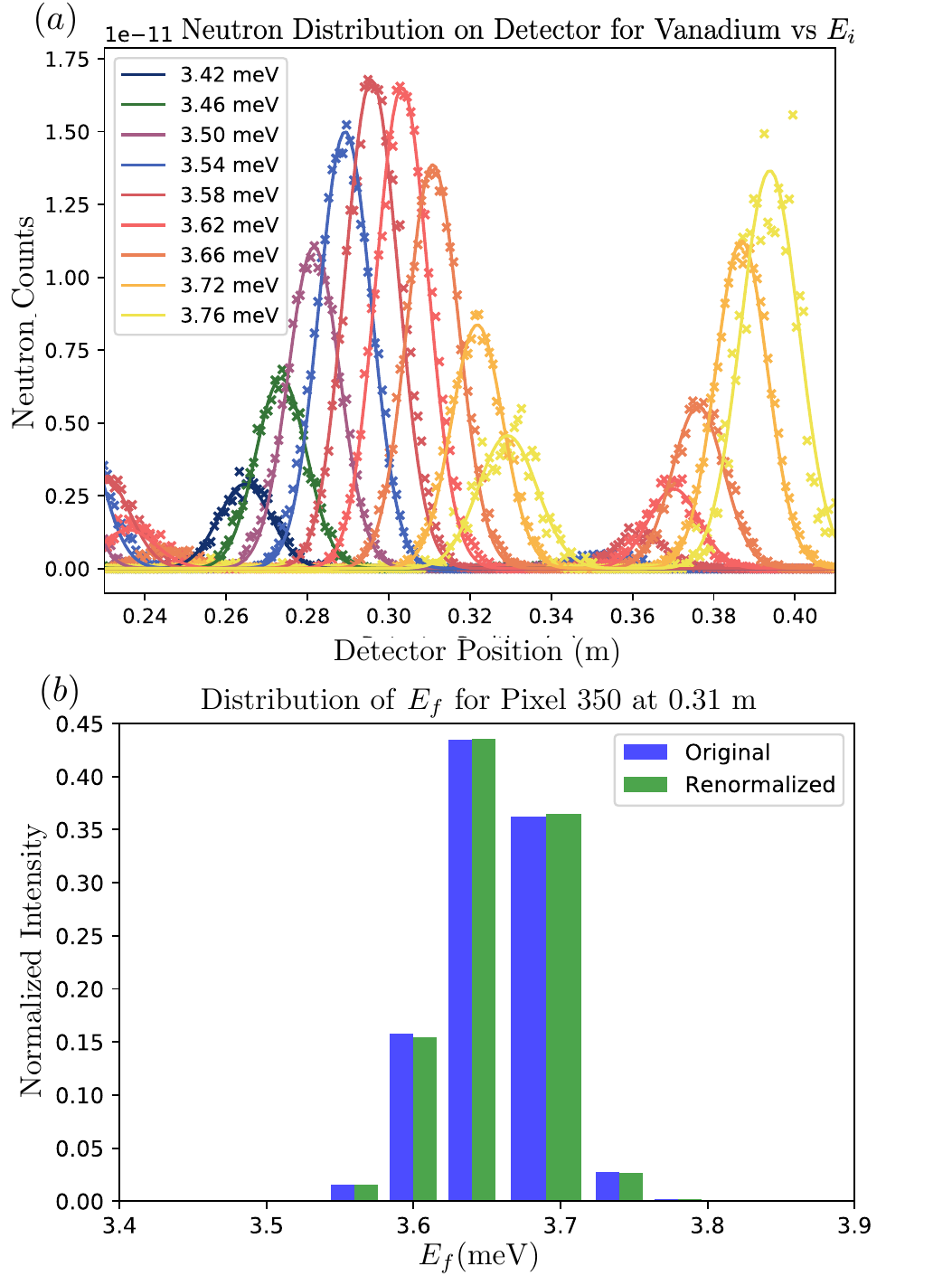}
    \caption{(a) Intensity versus detector position for nine representative calibration $E_i$'s scattered by analyzer crystals with a mosaic of $2^{\circ}$. This data was produced in an identical process to Fig.~\ref{fig:pcpa-calibration} but yields a visibly more intense signal for energies $E_i$ far from the nominal $E_f(s)$. (b) Histogram distribution representing the likelihood that a neutron detected on a given pixel had energy $E_f$. Note that the renormalization is less pronounced than in Fig. \ref{fig:histograms}, indicating that all $E_f$'s,  even those deviating from the Bragg condition, are intrinsically measured more homogeneously.}
    \label{fig:doublemosaic}
\end{figure}

 \begin{figure*}[t]
    \includegraphics[width=0.85\textwidth]{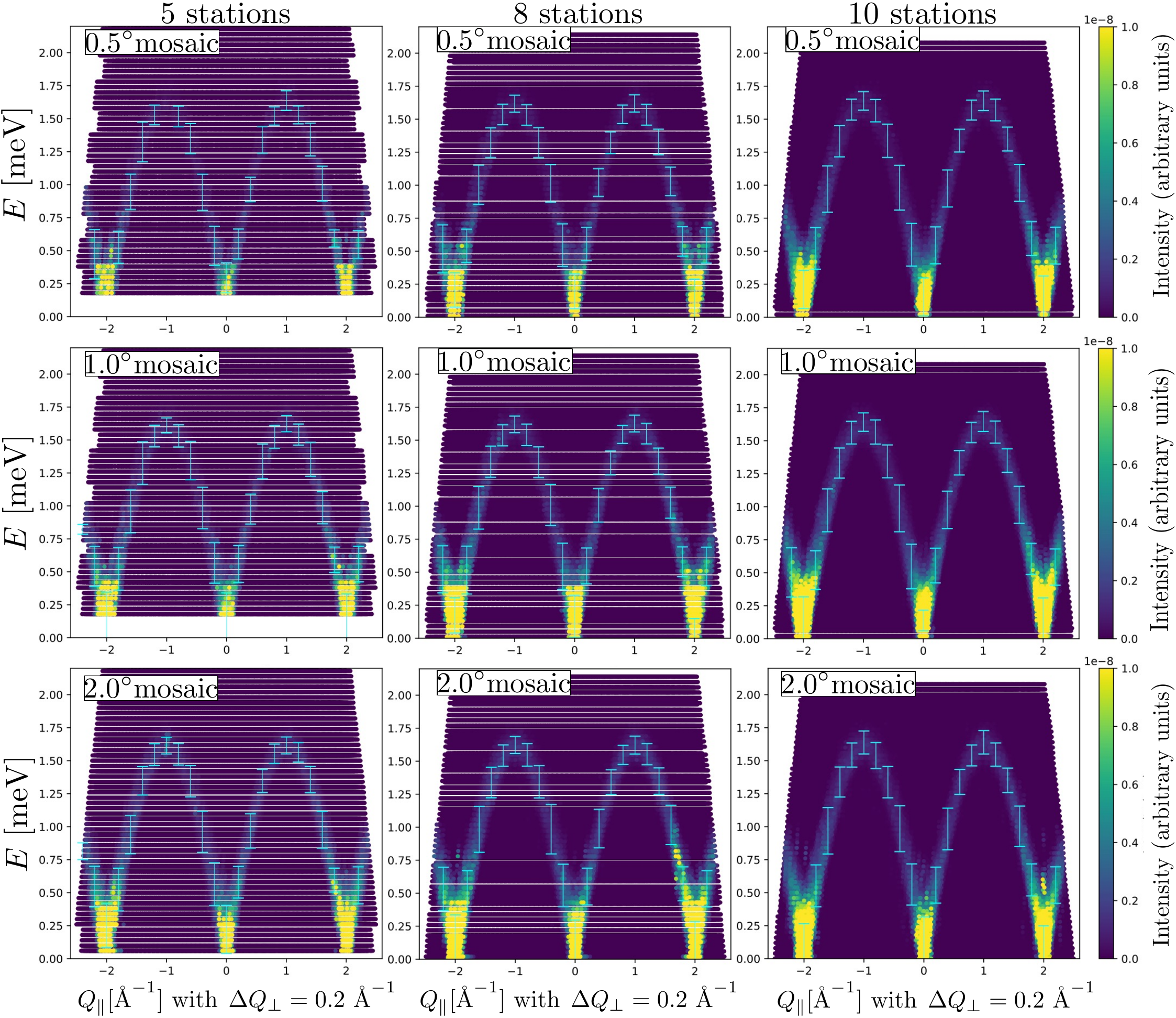}
     \caption{Scattering intensity for different secondary spectrometer designs and a simplified primary spectrometer, represented as momentum-energy slices with momentum transfer $Q_\parallel=Q_x$ for $Q_\perp = Q_y=-2.0$~\AA. The columns show the effect of increasing analyzer mosaic, starting from a $0.5^\circ$ mosaic on the top and ending with a $2^\circ$ mosaic at the bottom. Similarly, the rows show the effect of increasing the number of analyzer stations, beginning with 5 analyzer stations on the left and ending with 10 on the right. All designs were tested using 57 values of $2\theta$, 72 angles of $\phi$, and 2 $E_i$ values. Additionally, the width of the dispersion (cyan), calculated in the same manner as in Section \ref{fig:qyslices} is plotted on top of the dispersion. These results show that the PCPA approach is the most efficient with a coarse crystal mosaic because energies are measured more homogeneously. Additionally, more energies can be analyzed using  2$^\circ$ crystal mosaic analyzers, reducing gaps in energy in the data.}
     \label{fig:all_Designs}
 \end{figure*} 

To test this idea, we modified our virtual model to coarsen the analyzer mosaic to $2^\circ$ instead of $1^\circ$. The broader bandwidth of energies scattered by a single analyzer imply that the secondary spectrometer measures final energies more homogeneously, as seen from our calibration runs on Fig.~\ref{fig:doublemosaic}(a). As detailed in Sec.~\ref{sec:data analysis}, to produce pixel-energy histograms in Fig.~\ref{fig:doublemosaic}(b), we first re-scale the profiles of Fig.~\ref{fig:doublemosaic}(a) to account for intrinsically weaker scattering of specific energies. However, the scale factor will also correspondingly scale the signal uncertainty. Increased analyzer mosaic lowers the scale factor for most incident energies $E_i$, reducing the statistical error from scaling the Gaussian distribution. We note that several effects not accounted for in our simulations may limit the applicability of this concept. A $2^{\circ}$ mosaic  will require thicker crystals to account for the decrease in peak reflectivity, which will result in larger background effects from thermal diffuse scattering and inelastic phonon scattering from the crystals themselves \cite{HOPG, HOPG_Petersen}. This effect is complex and has not been modeled in our McStas simulations. The effect of $2^\circ$ and $0.5^\circ$ mosaic for our sample kernel simulations is shown in Fig.~\ref{fig:all_Designs}.

\subsection{Towards Continuous Energy Coverage}

With 8 analyzer stations, there is a gradual increase in the $E_f$ spacing of the analyzers, resulting in a non-continuous coverage of energy transfer. This increase in spacing with increasing $E_f$ is chosen due to the coarser energy resolution for the larger $E_f$ analyzers, thereby minimizing overlap between the energies scattered by differing stations. To eliminate gaps in the measured neutron scattering spectrum, an additional measurement, using, for instance, $E_i + 0.17$ meV, must be combined with the original $E_i$, which doubles the measurement time. However, with the PCPA method, it is, in principle, ideal to include overlap in the energies an analyzer will measure. The calibration accounts for, and in fact improves if multiple analyzers scatter neutrons with the same energy $E_f$. This is because the uncertainty from using the scale factor $G(E_f)$ for normalization is reduced. 

Using this guiding principle, we modified the virtual secondary spectrometer to include 10 analyzer stations, using constant steps in the analyzer's energy $E_f$, beginning at $E_f=$3.20 meV and increasing in intervals of 0.20 meV up to $E_f=5.0$~meV. The upper limit of $E_f=5$~meV is constrained by the need to use a beryllium filter to remove higher-order neutron wavelength contamination from the scattered beam, while the lower limit of $E_f=3.20$ meV was chosen due to increasingly large scattering angles required for low-energy measurements. To explore a lower cost alternative, we also created a 5 stations design with even intervals of 0.40~meV over the same $E_f$ range. By performing simulations with $E_i$ and $E_i + 0.20$ meV, it is possible to obtain a complete neutron scattering spectrum using the PCPA method.

To compare all three designs at different analyzer mosaic values, we employed a ``toy model" approach. The toy model follows a similar simulation procedure as described in Section \ref{sec:Methods} but with a few modifications to ease computational intensity. First, a planar, circular source with radius of 1 cm was focused towards a doubly focusing monochromator identical to the one planned to be used in the front end \cite{GRANROTH2024169440}. The source-to-monochromator and monochromator-to-sample distance was set to the exact same distance as the planned primary spectrometer, 1.6 meters. As a result, the spot size at the sample location using this simplified front end was nearly identical to the one produced by the full primary spectrometer. All of the designs required separate vanadium calibrations conducted in the same manner as described in Section \ref{sec:Methods}, except for a few modifications. First, the source was chosen to be monochromatic to have a more precise calibration for a given $E_f$. Additionally, for large analyzer mosaics, final energies that were previously negligible need to be accounted for, thus we used additional calibration energies as needed to properly interpolate the signal. While the simulations employing the vanadium sample for the calibration used an entirely monochromatic source term, the source used to produce the dispersion curve using the inelastic scattering kernel used a bandwidth of energies where ($\frac{\Delta E_i}{E_i} = 0.2$), which produced a similar energy resolution at the sample location as the full primary spectrometer. Finally, all designs used a second $E_i$ in addition to $E_i=5.1\unit{meV}$. The 5 station design used $E_i=5.3\unit{meV}$, the 8 station design again used $E_i=5.27\unit{meV}$, and the 10 station design used $E_i=5.2\unit{meV}$. The results, shown in Fig.~\ref{fig:all_Designs} for a typical momentum-energy slice, show that, in general, the integrated count rate increases with mosaic with little visible degradation in resolution. However, the 10 station $2^\circ$ mosaic analyzer design produces artifacts outside the signal dispersion. This may occur from the increased likelihood of scattering from multiple analyzers due to the proximity of the analyzers in the 10 station design and the large acceptance of the analyzer from the use of $2^\circ$. Additionally, note that these simulations do not consider the decrease in peak reflectivity, and therefore the increase in total count rate will be partially mitigated. As mentioned previously, the increase in mosaic is not a fully independent parameter and will result in other effects not fully modeled in McStas, such as background from thermal diffuse scattering.

\begin{figure}[h]
    \centering
    \includegraphics[width = 0.45\textwidth]{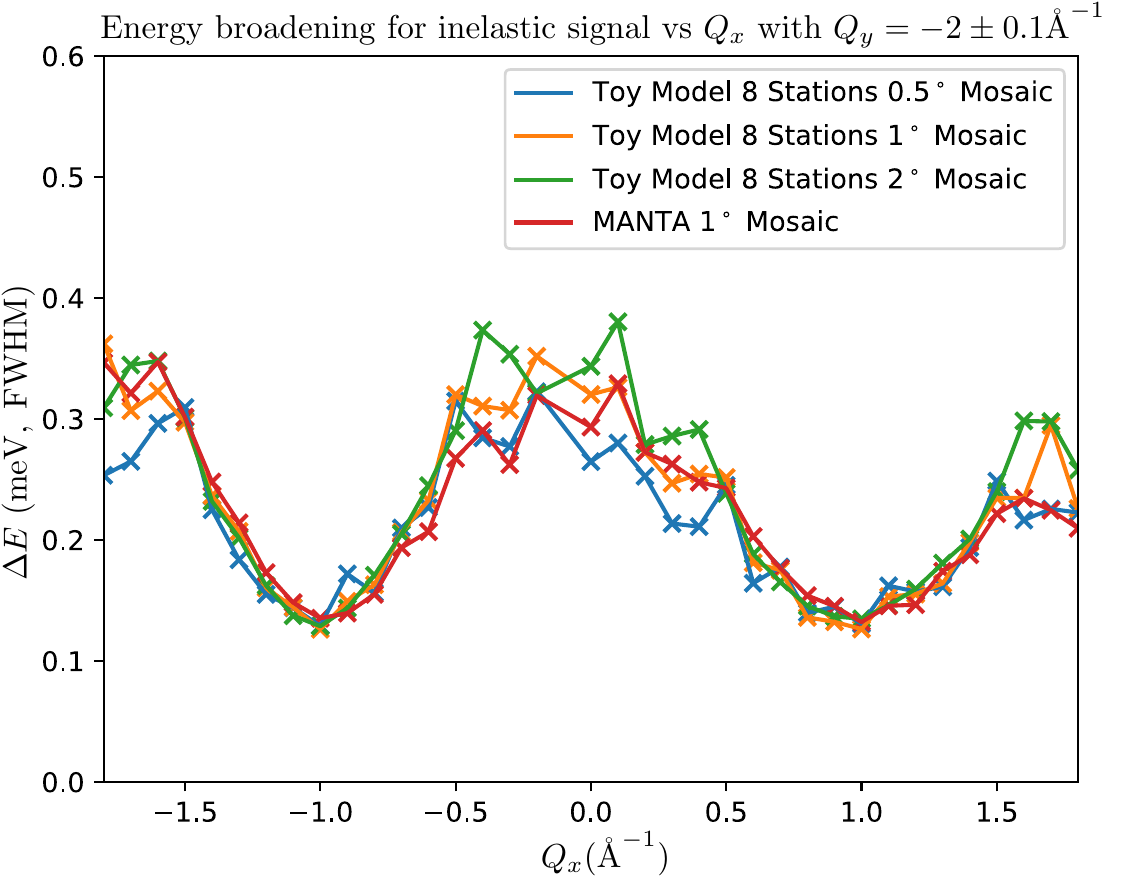}
    \caption{Effective energy broadening measured on a dispersive signal extracted from Figs.~\ref{fig:qyslices}(a) and \ref{fig:all_Designs} after fitting constant-$Q$ cuts in intervals of $Q_x = $ 0.1 $\unit{\AA^{-1}}$. The broadening is measured as the FWHM of the resulting Gaussian profile.}
    \label{fig:energy Resolution Phonon}
\end{figure}

To better quantify the quality of the results, we investigated the single crystal dispersion signal's energy broadening due to MANTA's finite energy resolution. We did this by fitting the data in Figs.~\ref{fig:qyslices}(a) and Fig.~\ref{fig:all_Designs} to constant-${Q}$ cuts separated by intervals of 0.1 $\unit{\AA^{-1}}$ and subsequently fitted with a Gaussian energy profile using $E$ bins with a width of 0.04~meV. The subsequent fitted Gaussian FWHMs as a function of $Q_x$ are shown in Fig.~\ref{fig:energy Resolution Phonon}. The effective energy broadening on the dispersive signal using the full MANTA spectrometer ranges from 0.2 to 0.35~meV, comparable to the resolution of the CTAX spectrometer \cite{CTAX} and CAMEA \cite{CAMEA_2023}. We also see that the 
energy broadening of the $1^\circ$ mosaic toy model and the full spectrometer are nearly identical, confirming the validity of the toy model approach. We also note that apart from the region near $Q_x=0$, the $2^\circ$ mosaic toy model produces identical energy broadening as the other three designs. Thus, using prismatic analysis, the dispersion energy resolution appears independent of crystal mosaic, which agrees with the results reported in the original prismatic concept paper \cite{Birk_prismatic}.  It is also worth noting that despite the potential background challenges from coarse mosaicity, using a mosaic as high as $2^\circ$ is not without precedent. The indirect time-of-flight spectrometer FARO \cite{FARO}, which also uses prismatic analyzers, decided to use this coarse mosaic value.

While the energy-broadening of the single crystal dispersion signal successfully depicts instrument-induced resolution effects, the $Q$-broadening of the single crystal are dominated by the sloped dispersion signal. The integration volume necessary to collect enough statistics ($\Delta Q_\parallel \approx 0.1 \unit{\AA^{-1}}$) is large enough where the dispersing signal is the largest contribution of the calculated full-width-half-maximum (FWHM). For this reason, an elastic Bragg scattering sample was used to calculate the elastic $Q$ broadening using the full primary and secondary spectrometer, as described in Section \ref{sec:Methods}. The intensity of the Bragg peak was integrated over a volume of $\Delta E = 0.04$ meV and $\Delta Q = 0.01 \unit{\AA^{-1}}$, and then the remaining $Q$ resolution was fit to a Gaussian and is tabulated in Tab. \ref{tab:q-res}. The broadening of both  $Q_\parallel$ and $Q_\perp$ (via $\theta$--$2\theta$ and transverse scans respectively) were calculated. Additionally, a single slit with an opening of $1^\circ$ was placed in front of the analyzer channel and compared to the setup with no slit used in the other simulations.

\begin{table}[]
    \centering
    \begin{tabular}{|l c c|}
    \hline
        & \textbf{\centering Elastic $Q$ broadening}  & \\ 
        \hline
          &  $\Delta Q_\perp$ $(\unit{\AA^{-1}})$ & $\Delta Q_\parallel$ $(\unit{\AA^{-1}})$ \\
        \hline
         $E_i=3.21$ meV  & 0.069 & 0.080\\
         \hline
         $E_i=3.21$ meV $1^\circ$ slit & 0.048 & 0.060\\
         \hline
         $E_i=4.05$ meV  & 0.051 & 0.075\\
         \hline
         $E_i=4.05$ meV $1^\circ$ slit & 0.034 & 0.045\\
         \hline
         $E_i=5.01$ meV & 0.050& 0.076\\
         \hline
         $E_i=5.01$ meV $1^\circ$ slit & 0.030 & 0.061\\
         \hline
    \end{tabular}
    \caption{Elastic $Q$ broadening calculated from an elastic Bragg scattering sample. The $Q$ broadening was calculated by integrating over a volume of $\Delta E= 0.04$ meV and $\Delta Q = 0.01 ~\unit{\AA^{-1}}$ in the orthogonal direction of the component of the $Q$ broadening being calculated. We calculated the broadening for both $Q_\perp$ and $Q_\parallel$ across multiple $E_i$ as well as in an open set up and one with a single slit with an opening of $1^\circ$.}
    \label{tab:q-res}
\end{table}

\section{\label{sec:Conclusion} Conclusion}

In this work, we have performed Monte-Carlo ray-tracing simulations of several multiplexed triple-axis spectrometer models relying on prismatic analysis to obtain a quasi-continuous energy-transfer coverage with only one or two incident energies $E_i$. Building on the concept and the design of the CAMEA spectrometer at PSI, we have introduced a distinct data analysis approach relying on the statistical likelihood that a neutron detected on a given pixel corresponds to a particular final energy $E_f$. This approach is dubbed Positionally-Calibrated Prismatic Analysis (PCPA). Simulations of the spectrometer model for a simple inelastic scattering signal show that PCPA data analysis maximizes the number of final energies measured simultaneously. Departing from the CAMEA design with eight analyzer stations, we explored further design evolutions, including changing the number of analyzer stations and the crystal mosaic of the graphite analyzers. These conceptual studies highlight the enormous flexibility offered by prismatic analysis. In principle, any existing TAS with a PSD located far enough from the analyzer to allow for distance collimation effects to occur can make use of the prismatic analyzer concept and the PCPA technique. Our conceptual work serves as a foundation for performing more comprehensive studies of multiplexed prismatic spectrometers, taking collimation and background mitigation into account. This includes testing a more complex sample kernel, considering thermal diffuse scattering from the analyzers, and dealing with complex sample environments. Our work utilized a realistic primary spectrometer developed in Ref.~\cite{GRANROTH2024169440}, including a fully optimized neutron guide system for the cold source of the High Flux Isotope Reactor. Although no formal concept has been chosen for the secondary spectrometer, when combined with a refined, CAMEA-style multiplexed prismatic system, a concept known as MANTA, our work shows that this modern beam-line would deliver world-class performance to study excitations in quantum condensed matter physics.
\begin{acknowledgments}
The work at Georgia Tech was supported by the Department of Energy, Basic Energy Sciences, Neutron Scattering Program under grant DE-SC-0018660. Initial work by Adit S. Desai was supported by the Georgia Institute of Technology's Letson Fellowship and the President's Undergraduate Research Award. The work of Marcus Daum was supported by the U.S. Department of Energy, Office of Science, Office of Workforce Development for Teachers and Scientists, Office of Science Graduate Student Research (SCGSR) program. The SCGSR program is administered by the Oak Ridge Institute for Science and Education for the DOE under contract number DE‐SC001466. This work used resources of the High Flux Isotope Reactor at Oak Ridge National Laboratory which is a DOE office of Science User Facility. We are grateful for the fruitful discussions with Mark Lumsden, Lowell Crow Jr., Jaime Fernandez-Baca, Barry L. Winn, Michael Hoffmann, Ian Turnbull, Mads Bertleson, Daniel Mazzone, and Jonas Birk that helped advance this work. We are indebted to Felix Groitl for providing us with the McStas model of CAMEA.
\end{acknowledgments}


%

\end{document}